\documentstyle[epsf]{l-aa}
\begin{document}
\def\fu{$f_1$}
\def\t{$\pm$}
\def\fd{$f_2$}
\def\fdu{$f_2 - 2f_1$}
\def\fp{$f_1 + f_2$}
\def\fm{$f_2 - f_1$}
\def\cd{cd$^{-1}$}
\def\cds{cd$^{-1}$\,}
\def\kms{km~s$^{-1}$}
\def\kmss{km~s$^{-1}$\,}
\def\I{\'\i}
\def\salp{\vskip 0.3truecm}
\thesaurus{6(03.13.2 - 08.15.1 - 08.22.1 - 10.19.2)}
\title{The galactic double--mode Cepheids}
\subtitle{I. Frequency analysis of the light curves and comparison with
single--mode Cepheids}
\author{I. Pardo\inst{1} and E. Poretti\inst{2}}
\institute {Universit\`a di Milano, I-20100 Milano, Italy
\and
Osservatorio Astronomico di Brera, Via Bianchi 46,
I-22055 Merate, Italy\\E--mail: poretti@merate.mi.astro.it}
\offprints{E. Poretti}
\date{Received date; Accepted Date}
\maketitle
\markboth{I. Pardo and E. Poretti: The galactic double--mode Cepheids I. The
frequency analysis}{ }
\begin{abstract}
We submitted the available photometric $V$ data of all the known galactic
Double Mode Cepheids (DMCs) to a careful frequency analysis with the aim of
detecting in each case the importance of the harmonics and of the cross coupling
terms. For each object, starting from different data subsets, we progressively
built a homogenous set of data, checking  the consistency of the results 
step by step. It was demonstrated that each star displays a different content,
showing that no a priori fit can be applied. Up to 4 harmonics were found for 
the fundamental radial mode ($F$); in every case, 2 harmonics were found for the
first overtone radial mode (1$O$). We also proceeded to a preliminar analysis
of the Fourier parameters of the DMC
light curves and we found a very close similarity between {\it i)} the light 
curves of the Classical Cepheids and those of the $F$--mode of the DMCs;
{\it ii)} the light curves of the $s$--Cepheids and those of the 1$O$--mode
of the DMCs. 

The analysis of DMC light curves offers the possibility of unifying the light 
curves of Classical and $s$--Cepheids. The case of the unique DMC CO Aur
is also discussed.

\end{abstract}
\section{Introduction}
The Double Mode Cepheids (DMCs) play an important role in the study
of the stellar evolution.
In the recent years a substantial improvement was made to reconcile
the pulsational mass (i.e. the mass predicted by the pulsation law 
$Q=P\sqrt{\rho}$), the beat mass (i.e. the mass derived from the ratio 
between the observed periods) and the evolutionary mass
(i.e. the mass predicted from  evolutionary tracks and luminosity). The 
introduction of new opacities allowed theoretical studies to fill not
only the large gap between the beat and pulsation masses, but also to match
the evolutionary masses (Christensen--Daalsgaard \& Petersen 1995).

In the same years, following the idea first expressed by Antonello et al.
(1990), Mantegazza \& Poretti (1992) and Poretti (1994) carefully studied
the light curves of {\it s}--Cepheids by using the Fourier
decomposition technique; they redefined the {\it s}--Cepheids as the stars
which do not follow the Hertzsprung progression (described by the
Classical Cepheids) in the space of Fourier parameters. To explain this
 different behaviour it was suggested that the two classes are pulsating in
 two different modes, i.e.
the fundamental radial ($F$) mode and the first overtone radial (1$O$) mode,
respectively. The DMCs provide the obvious
laboratory where verify this suggestion can be verified since it is a well established fact that
in 13 cases out of 14 the two excited modes are indeed the fundamental  and the
first overtone  mode; the data on V371 Per (Schmidt et al. 1995), the most
promising 15$^{\rm th}$ candidate, are too scanty to establish its DMC nature.
In the meantime, the large amount of data collected in the framework of the
MACHO (Alcock et al. 1995) and the EROS (Beaulieu et al. 1995) projects yielded
 the first confirmation of the different
pulsation modes since the Classical and {\it s}--Cepheids are separated in a
$P-L$ plane exactly by the shift due to the 1$O/F$ ratio. Moreover,
 new arguments were added to the debate owing to the large number
of DMCs discovered in the LMC, against the  only 14 cases observed in the
Galaxy.  To define in an accurate way the properties of the small number of
galactic DMCs is mandatory to perform a significant comparison with the 
properties of the more numerous LMC DMCs.

The light curve of a DMC can be considered as the sum of the contributions
of a number of frequencies, of which two only are independent (\fu and \fd).
Since each of these two curves is not, as a rule, perfectly sine--shaped, we
also have to observe the 2\fu, 3\fu, 4\fu, ..., 2\fd, 2\fd, 3\fd, 4\fd ...
harmonics; moreover, the two modes are interacting
and the cross coupling terms (i.e. their combination $|$~i\fu$\pm$j\fd~$|$;
the two cases \fm, \fp are the most frequent) are expected to be observed.
Even if systematic photoelectric surveys of DMCs were performed from 1947
onward (TU Cas; Oosterhoof 1959), no exhaustive
study of their light curves was carried out; the most complete analysis was
surely the one outlined by Stobie \& Balona (1979). However, in that
important paper also the light curve description was made on the basis of an
{\it a priori} choice, i.e. the application of a 2$^{\rm nd}$--order fit to the
collected points. This approach was also used by
Faulkner (1977) to study the light variation of U TrA: he applied three
different fits (3$^{\rm rd}$, 4$^{\rm th}$, 5$^{\rm th}$ order), but he
did not investigated whether all the components were really present in the data,
since the major result (i.e. the presence and the strength of the
cross--coupling terms) is slightly affected by the completeness of the
frequency content. Stobie \& Balona were mainly interested in the phasing
of the magnitude, colour and radial velocity observations and they showed,
in the particular case of VX Pup, that the effect of additional high--order
terms was to change only slightly the amplitude and the phases of the
low--order terms, not affecting their main result. We can conclude that {\it in
previous works  no attempt was done to detect how many harmonics of \fu and
\fd are
necessary to fit the observed light curves  and which cross coupling terms
are excited by their interaction}. More recently, this incomplete
approach was used by
Matthews et al. (1992) in reexamining the TU Cas data: a 4$^{\rm th}$--order fit
was {\it a priori} applied to the data, thus obtaining incorrect
values for the phase parameters and inconsistent amplitude ratios
(see also Poretti 1994 and Subsect. 4.4).

Therefore, it seems crucial to submit all the available photometry on DMCs
to a careful frequency analysis:
\begin{enumerate}
\item To detect the importance of the harmonics and of the cross coupling
terms for each star and to evaluate the similarities. Frequency and amplitude
variations can be investigated and the search for a third independent
periodicity ca be carried out;
\item To compare the values of
the low--order Fourier parameters with those of the galactic
single--mode Cepheids. This comparison will allow us to establish the 
communalities between the two
classes and to give an independent confirmation of the different pulsation mode
observed in single--mode Cepheids;
\item To establish the properties of the Fourier parameters by determining
boundary values in order to compare observed and theoretical light curves;
\item To search for the signature of resonances between modes in the
Fourier parameter progression.
\end{enumerate}
The first two items are discussed in this paper, the last two will be in
a successive paper (Poretti \& Pardo 1996; Paper II).

\section{Light curve content detection}
In our approach to the light curve analysis we did not select any arbitrary
order of
the fit, as in the above mentioned cases, but we searched for the terms
(which would be independent frequencies, harmonics or cross coupling terms) 
really constituting the DMC light curve. To do this, we used the
least--squares power spectrum method (Vanicek 1971) since it allows us to
detect one by one the constituents of the light curves. We currently apply
this algorithm to 
multiperiodic $\delta$ Sct stars and in the past we already  applied it
to the frequency analysis of the DMC CO Aur (Antonello et al. 1986) and
EW Sct (Figer et al. 1991). As a final step we fitted the observed magnitudes
by means of the formula
\begin{equation}
V(t)= V_o + \sum_z {A_z \cos [2\pi f_z  (t-T_o) +\phi_z ]}
\end{equation}
where $f_z$ is the generic frequency, which can be an independent frequency
(\fu and \fd), a harmonic or a cross coupling term. 

Let us discuss the methodology in detail, step by 
step, by anticipating the analysis of the measurements carried out by Berdnikov
on AS Cas, the latest DMC discovered. In the first power spectrum of Fig. 1 the
peak at \fu=0.3306 \cd and its alias structure (i.e. the 1--$f$, $f$+1, 2--$f$,
$f$+2, ... terms) introduced by the spectral window are clearly visible.
The aliases are particularly strong in this dataset since the measurements were
obtained in a single site; when merging measurements obtained at two or more
sites the height of the aliases will decrease. Then we introduced \fu as
a known constituent searching for the second term: in the second power
spectrum the \fd=0.4639 \cd term and its alias structure appeared. It is
important to note that no prewhitening was done: only the frequency value
\fu was considered as established
(known constituent; k.c.) and in the second search the unknowns were $V_o,
A_1, \phi_1, f_2, A_2, \phi_2$. Before proceeding further with a new
frequency search,
the values of  \fu and \fd were refined by a simultaneous least--squares
fit and then they were introduced as k.c. in the third search, which allowed
us to detect the \fp term (third panel). Now, frequency refinement is a
delicate step because the third component must always satisfy the relationship
\fp; to do this refinement, we use the MTRAP code (Carpino et al. 1987) which
keeps this relationship locked throughout the best fit search. After the
refinement, we introduced the \fu, \fd, \fp terms as k.c. 
($V_o, A_1, \phi_1, A_2, \phi_2, A_{f_1+f_2}, \phi_{f_1+f_2}, f_3, A_3, \phi_3$
are the unknowns) searching for the new light curve component: we detected
2\fu. Once again, the refinement was performed by keeping the \fp and 2\fu
relationships locked; new frequency values were then obtained and 
introduced as k.c., the fifth component \fm was detected and so on. Following
this process, we detected 11 terms and we noted that in this latter case, the
highest peak is not the  2\fd--\fu term, but its alias at 1 \cd (see lower
left panel of Fig. 1). This overtaking is due to the interaction between 
noise (the terms
have an amplitude of only 11 mmag against a standard deviation of 26 mmag
for the measurements) and spectral window (we were dealing here with 
single--site measurements).  When observing this event, the exact value of
the cross coupling term is considered to proceed further.

 The decision to stop the term selection
was taken when no more term was visible over the noise distribution, i.e.
when all the terms giving a significant contribution to the light curve shape
were presumably identified. In Fig. 1 the 12$^{th}$ panel clearly shows that
no other term can be detected in a clear way as the noise distribution
is quite uniform. Of course, very small amplitude terms can remain hidden in
the noise level, especially when dealing with inaccurate measurements.
\begin{figure*}
\epsfxsize=15cm
\epsfysize=19.7cm
\centerline{\epsffile{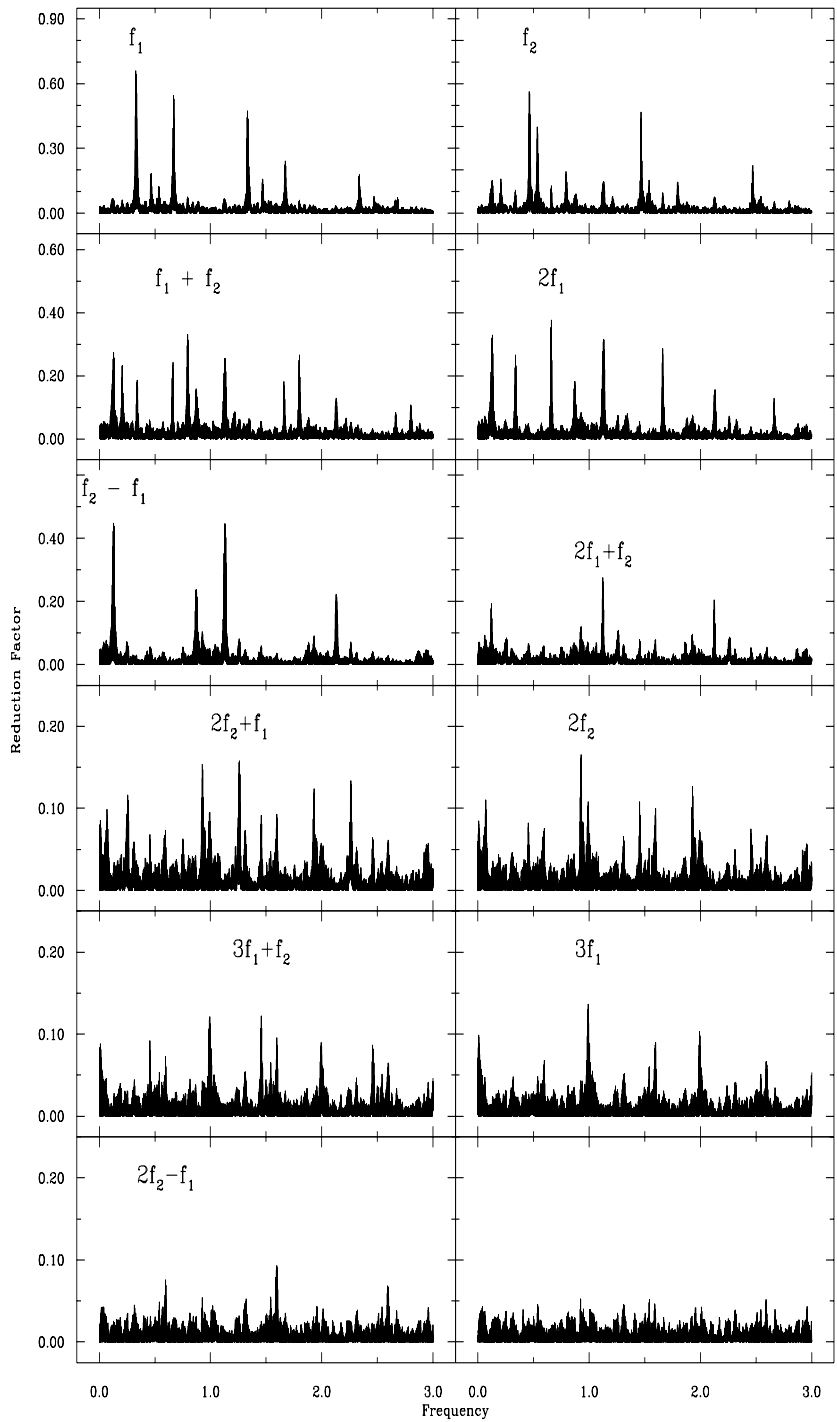}}

\caption[]{Power spectra of the AS Cas measurements. Each panel shows the
 spectrum obtained by introducing all the terms identified as k.c. in the
 previous ones: this means that their frequencies are considered as 
 established, but their amplitudes and phase values are recalculated for
 each trial frequency}
 \end{figure*}

In the frequency analysis it also occurred, at times, that after the 
detection of
the two frequencies, their harmonics and some coupling terms, the highest peak
in the power spectrum was at (or very close to) 1.003 \cd. Two preliminary
checks were performed: the first, quite obvious, step was to check that such
a term would not be an exotic 
coupling term (or any of its aliases); the next, to check that  the 1.003
\cd peak would indeed be an alias of a 0.003 \cd term. In the latter hypothesis, the peak
originated from a misalignment between measurements in different subsets
(or in different years within the same subset).
If these two checks were unsuccessful, the fact that
1.003 \cd means 1 sidereal day/day  suggested
the reasonable hypothesis that it was a spurious term introduced by an
instrumental and/or a methodological effect.
If the frequency analysis was not terminated, the 1.003 \cd
term was introduced as a known constituent, but no scientific meaning was 
attributed to it. 

As an example, let us consider the BERD measurements performed on BQ Ser,
obtained in different years; Tab. 1 lists  the
mean magnitudes for each year and systematic shifts are indeed observed.
The importance of misalignments 
is emphasized by Fig. 2, where the final power spectrum (i.e. the spectrum 
obtained by processing the BERD measurements considering \fu, \fd, 2\fu, \fp,
\fm, 2\fd, 2\fu+\fd, 3\fu  as k.c.) is shown 
with and without systematic corrections: the amplitude of the peak
at 0.0025 \cd in the upper panel corresponds to an amplitude of 0.014 mag;
after the end--to--end alignment (just before the final fit) this peak has
completely disappeared (lower panel).
\begin{table}
\begin{flushleft}
\caption{ As an example of  mean magnitudes differing from one year to the
next the case of the BERD measurements of BQ Ser is shown; the errors on each
value are a few mmag. Owing to the small number of
measurements in some years, a reliable determination of the mean magnitude
can be done only in the last steps  of the analysis}
\begin{tabular}{l c  rr}
\hline
\multicolumn{1}{c}{Year}& &\multicolumn{1}{c}{$V_0$} & \multicolumn{1}{c}{N}\\
\noalign{\smallskip}
\hline
\hline
\noalign{\smallskip}
1986 & & 9.500 & 25 \\
1988 & & 9.499 & 29 \\
1989 & & 9.498 & 40 \\
1990 & & 9.500 & 20 \\
1991 & & 9.506 & 31 \\
1992 & & 9.515 & 86 \\
1993 & & 9.525 & 51 \\
1994 & & 9.517 & 85 \\
\noalign{\smallskip}
\hline
All  & & 9.511 & 367 \\
\noalign{\smallskip}
\hline
\hline
\end{tabular}
\end{flushleft}
\end{table} 
\begin{figure}
\epsfxsize=8.8cm
\epsfysize=11.0cm
\centerline{\epsffile{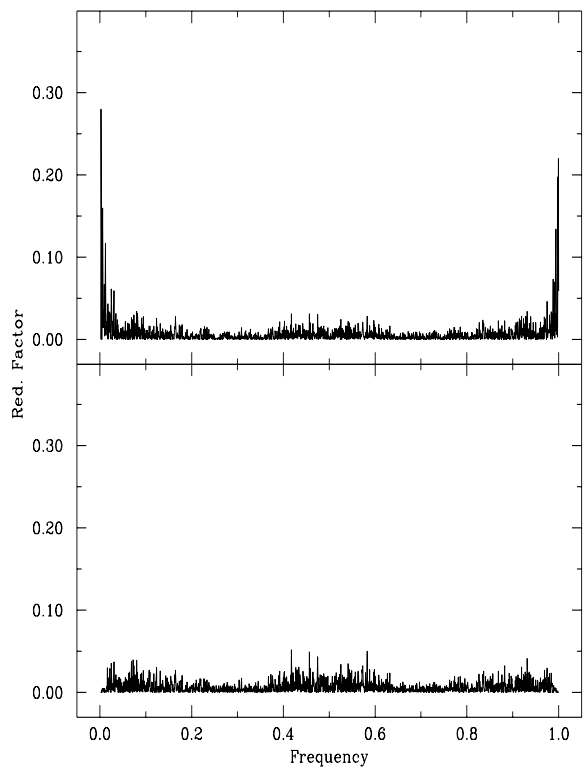}}
\caption[]{Effects of the year-to-year misalignments in the BERD
measurements of BQ Ser (upper panel: no correction; lower panel: after
application of the magnitude shifts): the peaks at 0.0 and 1.0 \cd
are strongly enhanced in the upper panel}
\end{figure}

\begin{table*}
\begin{flushleft}
\caption{Double--mode Cepheid stars. AM84: Antonello \& Mantegazza 1984;
AM86: Antonello et al. 1986; BB87: Babel \& Burki 1987; BERD: Berdnikov 1992,
Berdnikov et al. 1995abc;
B83: Barrell 1983; FP91: Figer et al. 1991; FS79: Faulkner \& Shobbrook 1979;
JA62: Jansen 1962; LJ65: Leotta Janin 1965; LTPV: Manfroid et al. 1991, 1994,
 Sterken et al. 1993;
 MI64: Mitchell et al. 1964; MS78: Madore et al. 1984; MV75: Madore \& Van
 den Berg  1975;  MB84: Moffett \& Barnes 1984; OO57:
Oosterhoof 1957; OO59: Oosterhoof 1959; P76: Pel 1976; SB: Stobie \& Balona
1979; WE57: Worley \& Eggen 1957}
\begin{tabular}{ll c lll c r c  rrr l}
\hline
\multicolumn{2}{c}{Star} & & \multicolumn{1}{c}{$f_1$} &
 \multicolumn{1}{c}{$f_2$}&\multicolumn{1}{c}{$f_1/f_2$} & &\multicolumn{1}{c}{N$_{\rm meas}$} & &
 \multicolumn{1}{c}{N$_{f_1}$}&\multicolumn{1}{c}{N$_{f_2}$}&
 \multicolumn{1}{c}{N$_{cc}$}&
 \multicolumn{1}{c}{References}\\
\hline
\noalign{\smallskip}
TU      & Cas & & 0.467442  & 0.658635 & 0.7097 & &618& & 4 & 2 & 9 &  
 OO59 WE57 BERD \\
U       & TrA & & 0.389344 & 0.547983& 0.7105 & & 1060& & 4 & 2 &  6 &  OO57 MI64 JA62  FS79 SB79 BERD \\
VX      & Pup & & 0.332030& 0.467384& 0.7104 & & 234 & & 3 & 2 &  6 &  SB79 MB84 LTPV BERD\\
AS      & Cas & & 0.330628& 0.463936&0.7127 & & 575 & & 3 & 2 &  6 &   BERD \\
AP      & Vel & & 0.319717& 0.454587&0.7033 & & 255 & & 3 & 2 &  5 &   P76 SB79 BERD \\
BK      & Cen & &0.315072 & 0.449860&0.7004 & & 251 & & 3 & 2 &  5 &  
 LJ65 LTPV SB79 BERD \\
UZ      & Cen & &0.299910 & 0.424589&0.7064 & & 131 & & 4 & 2 &  5 &   
 P76 SB79 BERD \\
Y       & Car & & 0.274742& 0.390698&0.7032 & & 137 & & 3 & 2 &  5 &  
 P76 SB 79 BERD\\
AX      & Vel & & 0.272241& 0.385657&0.7059 & & 520 & & 2 & 2 &  3 &  
 BA83 P76 SB79 BERD \\
GZ      & Car & & 0.240448& 0.340857&0.7054 & & 118 & & 2 & 2 &  3 &  
 P76 SB79  BERD\\
BQ      & Ser & & 0.234138& 0.331997&0.7052 & & 602 & & 3 & 2 &  3 &  
 MB84 LTPV BERD\\
EW      & Sct & & 0.171719& 0.245820&0.6986 & & 515 & & 3 & 2 &  3 &  
 FP91 BERD \\
V367    & Sct & & 0.158902& 0.228061&0.6968 & & 514 & & 2 & 3 &  2 &  
 MS78 MV75 BERD\\
\noalign{\smallskip}
CO      & Aur & & 0.560844& 0.700390&0.8008 & & 370 & & 3 & 1 &  2 &  
AM84 AM86 BB87 BERD\\
\noalign{\smallskip}
\hline
\end{tabular}
\end{flushleft}
\end{table*} 
\section{Data collection and reduction}
A bibliographic search for published measurements was carried out over
a very long time baseline; we thus found photometric data collected in
a variety of systems and using both absolute ({\it all--sky}) and
differential photometry. Since involved 
amplitudes are large and vary in function of the wavelength,
it was considered necessary to restrict the analysis to a well defined
passband. The choice of the $V$ filter of the $UBV(RI)$ system was  
quite natural since this filter was by far the most used; the $y$ filter
in the $uvby$ photometric system was considered as equivalent. We also 
considered measurements carried out in other photometric systems only if
the authors themselves supplied a  transformation formula from his own system
to the $UBV$ one. However, the actual consistency between the $\lambda_{\rm eq}$ of
photometric systems having different passbands is difficult to admit and also
small differences can seriously  affect the observed amplitudes (see the
case of AX Vel).

Moreover, even if the photometric system is the same, the mean magnitude of
the light variation is expected to be different. When absolute (or standard)
photometry is performed, systematic differences of a few hundredths of 
magnitude are common in the transformation from instrumental to standard
system. When differential photometry was performed, different
values for the magnitude of the comparison stars were used, creating
once again systematic differences. 

Another serious problem was the large gaps between the different data subsets
for the same star. If several years elapsed without any measurement,
the data analysis was not a simple task since, for example,
periods or amplitudes can change; also time series analysis are much more
time--consuming (we adopted a frequency step of 1/10$\Delta$t, where $\Delta$t
is the difference between the times of the last and first measurements).
\smallskip

This being considered, we applied the following procedure to the collected
measurements (the last column of Tab.2 reports the list of references):
\begin{enumerate}
\item We performed a very preliminary frequency analysis 
using the measurements reported by each author (in our terminology they 
constitute a {\it subset}), thus obtaining mean magnitude values;
\item It should be noted that a good spectral window, not only a satisfactory
number of points, was necessary for our purposes.
  If the gaps in time were not too
large, we merged two or more subsets, obtaining one or more {\it datasets};
to do this, we applied systematic shifts to align the subsets to the same mean
magnitude level.  Hence, these datasets were subject to
separate frequency analysis and 
the components of the respective light curves were detected. In turn, the
parameters of the least--squares fit were calculated;
\item In principle, the small amplitude terms detected in one dataset were
not the same as the ones detected in the others and a major difficulty was to
understand which
of them should be considered as real. Indeed, small amplitude terms were
strongly affected by the noise, different from one dataset to the
other; as a result, some terms could stand out in the frequency analysis of
a dataset and remain hidden in the noise of another.
How could we decide which terms had to be used to describe the light curve of 
a DMC? To that purpose,
we included the frequency of a component clearly evidenced in a
dataset among the input values of a least--squares fit ({\it a forced fit})
of another dataset, not showing it. Therefore, we compared
the phase values: if  the datasets yielded similar values,
the component was considered to be significant, if not it was rejected;
\item On the basis of the previous results, all the datasets of a given DMC
(and also, in some case, the subsets we could not use for the frequency
 analysis) 
were merged into the {\it whole set} of data, and the final least--squares 
fit of the data was performed, together with 
additional tests. In particular, we checked again the mean magnitude levels
of each subset and we performed some further, minor adjustments. We also
performed a frequency refinement and we reported the values we obtained
in Tab. 2; the formal errors on the frequency values are of the order of 
a few 10$^{-6}$ \cd;
\item Once the frequency content was determined, we obtained the light
curves of the two frequencies \fu and \fd by subtracting the theoretical 
contribution of the other terms from the measurements. To obtain
the light curve on \fu we subtracted \fd, 2\fd, 3\fd, ...  and all the
coupling terms; to obtain the light curve on \fd we subtracted \fu, 2\fu,
3\fu, ...  and again all the coupling terms.
\end{enumerate}
In the next section we present a detailed description of this process as
applied to the  BQ Ser and AX Vel measurements.
It is important to note that this process led us to perform a final fit 
by using all the terms detected in the power spectra and these terms only:
{\it no a priori choice of the fit order was considered}.

To increase our confidence in the results, we also considered a different
final step: before merging the datasets
into the whole set of data, we determined the generalized
phase differences for each dataset and then we calculated their weighted mean
(this approach is fully described in Pardo 1995).
The comparison between these weighted values and the ones obtained from the
whole set of data showed that the two procedures yield equivalent results.
Moreover, it should be also noted that from
a chronological point of view, we progressively built up each set and 
performed several preliminar frequency analyses and fits: the stability of 
the amplitudes and phases of the constituents previously considered as well
established did not change appreciably. 
 As an example, we concluded the analysis before Berdnikov
reported a new series of photoelectric data (Berdnikov et al. 1995a, 1995b;
Berdnikov \& Turner 1995a, 1995b), but
we felt us obliged to revise the results obtained so far (reported in Pardo 
1995). This further extension of the available data
strengthened our confidence since it did not 
produce any significant numerical change; the only, but remarkable, 
exception was to give stronger evidence of the 2\fd and 3\fu terms in the light curve
of EW Sct.

The whole sets of data used for the analysis can be requested from the authors.

\section{Star by star}
In this  section we shortly review each star, reporting a detailed
description for BQ Ser and AX Vel only; a thorough discussion of
the frequency and least--squares analysis of all the DMCs was performed
by Pardo (1995). In the discussion we used the phase differences $\phi_{21}=
\phi_2 - 2\phi_1$ and $\phi_{31}= \phi_3 - 3\phi_1$ (which can be calculated
for both the \fu and \fd terms); 
the discussion of the {\it generalized phase differences} will take place in
Paper II. Table 2 summarizes the general results obtained on all the DMCs.
We report below (Subsect. 4.1) the full application of our procedure to the
BQ Ser measurements and Tab. 3 shows the results obtained step by step.
Tables 4 and 5 list the Fourier coefficients of the fits of the whole sets
of data for the 14 other stars.
\begin{table*}
\begin{flushleft}
\caption{The complete analysis of BQ Ser data. The MB84, BERD, LTPV subsets
were analyzed separately (upper panels); then the last two were merged into
the LTPV+BERD dataset (middle panel in the lower part). A fit was forced on the
MB84 data to check the phase values of undetected terms (left panel in
the lower part); since the check was positive (see text to compare phase
difference values), all the terms were included in the global fit (right panel
in the lower part)}
\begin{tabular}{l | rrr | rrr | rrr}
\hline
\multicolumn{1}{c}{Term}  &  \multicolumn{1}{|c}{Frequency} & \multicolumn{1}{c}{Ampl.} & \multicolumn{1}{c|}{Phase}  &
                             \multicolumn{1}{|c}{Frequency} & \multicolumn{1}{c}{Ampl.} & \multicolumn{1}{c|}{Phase}  &
                             \multicolumn{1}{c}{Frequency} & \multicolumn{1}{c}{Ampl.} & \multicolumn{1}{c}{Phase} \\
\multicolumn{1}{c}{} &  \multicolumn{1}{|c}{[cd$^{-1}$]} & \multicolumn{1}{c}{[mmag]} & \multicolumn{1}{c|}{[rad~10$^{-2}$]} & 
                         \multicolumn{1}{c}{[cd$^{-1}$]} & \multicolumn{1}{c}{[mmag]} & \multicolumn{1}{c|}{[rad~10$^{-2}$]} & 
                         \multicolumn{1}{c}{[cd$^{-1}$]} & \multicolumn{1}{c}{[mmag]} & \multicolumn{1}{c}{[rad~10$^{-2}$]} \\
\noalign{\smallskip}
\hline
\hline
\noalign{\smallskip}
     & \multicolumn{3}{c}{Subset: MB84} & \multicolumn{3}{c}{Subset: LTPV} & \multicolumn{3}{c}{Subset: BERD}\\
\noalign{\smallskip}
\hline
\noalign{\smallskip}
 \fu       & 0.234140 & 176\t3 & 56\t2 &  0.234143 & 179\t2 &  155\t2 &  0.234136 & 176\t1 & 157\t1 \\  
 \fd       & 0.332020 & 109\t2 & 109\t2 &  0.331985 & 112\t3 &  232\t2 &  0.331999 & 111\t1 & 236\t1 \\  
2\fu       &          & 34\t3 & 535\t9 &           & 29\t3 &  91\t10 &           & 31\t1 & 110\t4 \\
\fu+\fd    &          & 38\t2 & 618\t9 &           & 36\t3 &  210\t8 &           & 36\t1 & 214\t4 \\
\fd--\fu    &          & 18\t2 & 526\t12 &           & 20\t3 &  535\t14 &           & 20\t1 & 532\t7 \\
2\fd       &          &       &      &           &(5\t2)& (328\t64) &           & 5\t1 & 364\t26 \\
3\fu       &          &       &      &           & 6\t3 &  19\t49 &           & 6\t1 & 75\t26 \\         
2\fu+\fd   &          & 12\t3 & 446\t25 &           & 9\t2 &  168\t34 &           & 7\t1 & 154\t20 \\
\noalign{\smallskip}
$V_0$     & \multicolumn{3}{|c|}{9.5137$\pm$0.0015} & \multicolumn{3}{|c|}{9.5139$\pm$0.0015} & \multicolumn{3}{|c}{9.5110$\pm$0.0009}\\
rms       & \multicolumn{3}{|c|}{0.0126 mag} & \multicolumn{3}{|c|}{0.0157 mag} & \multicolumn{3}{|c}{0.0169 mag}\\
N         & \multicolumn{3}{|c|}{81} & \multicolumn{3}{|c|}{121} & \multicolumn{3}{|c}{367}\\
$T_0$     & \multicolumn{3}{|c|}{HJD 2444397.3965} & \multicolumn{3}{|c|}{HJD 2448199.2453} & \multicolumn{3}{|c}{HJD 2448199.2453}\\
\noalign{\smallskip}
\hline
\hline
\noalign{\smallskip}
     & \multicolumn{3}{|c}{Dataset: MB84} & \multicolumn{3}{c|}{Dataset: LTPV + BERD} & \multicolumn{3}{|c}{Whole set}\\
\noalign{\smallskip}
\hline
\noalign{\smallskip}
 \fu       & 0.234140 & 174\t3 & 55\t2 &            0.234137 & 177\t1 & 156\t1    &     0.234138 & 177\t1 & 155\t1 \\
 \fd       & 0.332020 & 112\t4 & 110\t3 &            0.331999 & 112\t1 & 235\t1    &    0.331997  & 112\t1 & 236\t1 \\
2\fu       &          & 32\t3 & 536\t10 &                     & 30\t1 & 105\t4    &              & 31\t1 & 103\t3 \\
\fu+\fd    &          & 38\t3 & 612\t10 &                     & 36\t1 & 212\t3    &              & 36\t1 & 213\t2 \\
\fd--\fu    &          & 19\t2 & 524\t12 &                     & 20\t1 & 534\t6    &              & 20\t1 & 537\t4 \\
2\fd       &          &(4\t4)&(90\t93)&                     & 5\t1 & 363\t23    &              & 5\t1 & 367\t19 \\
3\fu       &          &(3\t5)&(436\t168)&                    & 6\t1 & 62\t18    &              & 6\t1 & 67\t16 \\
2\fu+\fd   &          & 11\t4 & 452\t31 &                     & 7\t1 & 157\t15    &              & 9\t1 & 152\t10 \\
\noalign{\smallskip}
$V_0$     & \multicolumn{3}{|c|}{9.5133$\pm$0.0016}& \multicolumn{3}{|c|}{9.5110$\pm$0.0008}& \multicolumn{3}{|c}{9.5107$\pm$0.0006}\\
rms       & \multicolumn{3}{|c|}{0.0125 mag}&\multicolumn{3}{|c|}{0.0167 mag}& \multicolumn{3}{|c}{0.0149 mag} \\
N         & \multicolumn{3}{|c|}{81}&  \multicolumn{3}{|c|}{488}& \multicolumn{3}{|c}{602}\\
$T_0$     & \multicolumn{3}{|c|}{HJD 2444397.3965} & \multicolumn{3}{|c|}{HJD 2448199.2453}&\multicolumn{3}{|c}{HJD 2448199.2453}\\
\noalign{\smallskip}
\hline
\end{tabular}
\end{flushleft}
\end{table*} 
\subsection{BQ Ser}
 The full analysis of this star is described in Tab. 3; let us examine 
 it in detail.
The preliminary analyses of the three subsets yielded some slight differences
in the  frequency content. Firstly, the BERD subset (upper part, right panel)
 evidenced the 2\fd term,
but this term was not detected by the frequency analysis of the other subsets; 
however, a forced
fit on the LTPV data (upper part, middle panel) yield a very similar phase
value (between brackets;
note that the $T_0$'s are the same) and this term was considered as real.
Then, the LTPV and BERD subsets were merged into one dataset (the measurements
were performed in the same years) and its frequency analysis evidenced once again
the same terms (lower part, middle panel), thus
confirming their agreement in phase.
 Now, we can compare the LTPV+BERD dataset with the MB84 one, where
the  2\fd and 3\fu terms were not detected (upper part, left panel). 
When these terms were added in the forced least--squares fit of the MB84 data,
 the parameters of the fit did not change appreciably
(compare the upper and lowel right panels of Tab. 3) and we obtained Fourier
parameters ($\phi_{21}$=4.98$\pm$0.95 rad for the \fd term
and $\phi_{31}$=2.71$\pm$1.70 rad for the \fu term) very similar to those
obtained by fitting the LTPV+BERD data (5.26$\pm$0.19 rad and 2.35$\pm$0.18
rad, respectively).
Considering all the terms, the maximum difference between the
amplitudes is 0.004 mag, within the formal error bars. Therefore we can
conclude that the light curve
of BQ Ser contains the \fu, \fd, 2\fu, \fp, \fm, 2\fd, 3\fu, 2\fu+\fd
terms. The right panel of the lower part of Tab. 3 lists the 
parameters of the final fit on all the available data (the
BERD measurements obtained by using a different equipment (Berdnikov \& 
Turner 1995a, 1995b) were added to the previous ones). It is useful to
verify once again that the phase and amplitude values of the final fit
are very similar to those of the fit of the BERD subset, as should happen
when  adding measurements with the same frequency content. 
The \fu and \fd values are practically coincident (over a 11--year basis).

Figure 3 shows the light curves of the two periods of BQ Ser: the upper
curve was obtained by subtracting the \fd, 2\fd, \fu+\fd, \fd-\fu, 2\fu+\fd 
terms from the measurements (whole set of data), the lower one by
subtracting \fu, 2\fu, 3\fu, \fu+\fd, \fd-\fu, 2\fu+\fd.
 To fit the upper
curve two harmonics are necessary, while the first harmonic is sufficient
to fit the lower curve. If the order of the fit had been established
{\it a priori}, the same harmonic contents would have been used, but this
assumption was not justified by the different shapes of the two light
curves.

Szabados (1993) claimed evidence for a third periodicity ($f_3$=0.42 \cd) in
the light curve of BQ Ser on the basis of his unpublished data. However, it
must be noted that the very common 2\fd harmonic was not detected and that its
value (0.66 \cd) is very close to one of the largest amplitude terms involving
the $f_3$ term (0.65 \cd). Hence, in our opinion, the $f_3$ term detection 
arose from a misidentification of the 2\fd component in an early step of the
frequency analysis; it should be noted that a plausible explanation for the
0.42 \cd term itself is the identification of the 2\fd--\fu term (0.43 \cd).

\begin{figure}
\epsfxsize=8.8truecm
\epsfysize=12.0truecm
\centerline{\epsffile{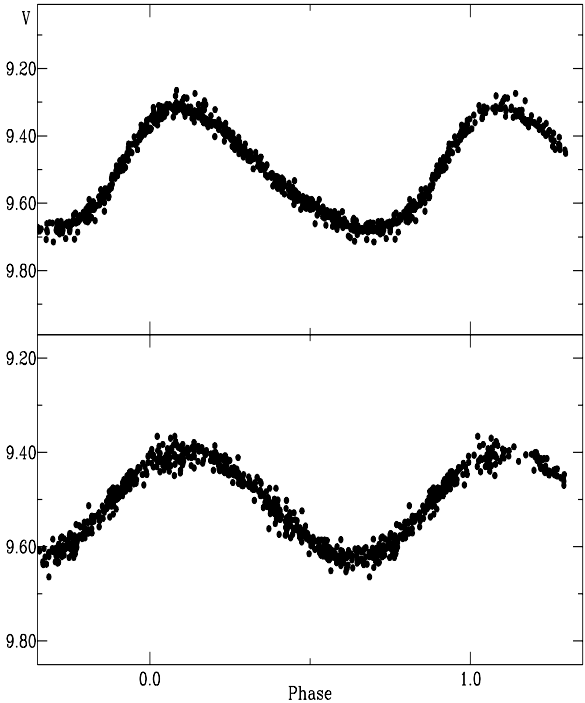}}

\caption[ ]{Light curves of the two independent frequencies \fu=
0.234138 \cd (upper panel) and \fd=0.331997 \cd (lower panel) as obtained
from the whole BQ Ser set of data}
\end{figure}
\subsection{AX Vel}
As a further example of the procedure reported above, the analysis of AX Vel
allows us to give a better description of  some other aspect.
We preliminarly scrutinized the three available subsets (BA83,
P76, SB79), calculating the mean magnitudes of each
of them (8.213, 8.219, 8.215, respectively) and then we merged the last two 
into one (PSB dataset). By performing the careful frequency analysis 
described in section 2, in both datasets we detected the \fu, \fd, \fm, \fp,
2\fu, 2\fd terms.
The frequency analysis of BA83 dataset also evidenced the 
2\fu + \fd terms, while that of the PBS dataset evidenced the 3\fu term.
To check if these two terms have a physical meaning, we
applied a least--squares fit separately to both datasets by using all the above
quoted terms. As a result, in the PSB dataset the 2\fu+\fd term had
a phase value of 3.1$\pm$0.3 rad (against 2.6$\pm$0.3 rad obtained in the
Barrell dataset); since the two values are similar and error bars
are overlapping, this  term is included in the light curve content of
AX Vel. On the other hand, the fit of the BA83 subset considering also the
3\fu~term yielded a large error bar on the phase value of this term 
(i.e. $\pm$ 2.6 rad), preventing a reliable confirmation; hence, this term
was dropped from the final least--squares solution.
When this analysis was concluded, Berdnikov \& Turner (1995) reported
on new measurements (26): they were not sufficient to perform a reliable 
frequency analysis, but 
by adding this subset we improved the frequency values (the time baseline 
being much longer) and then we included it in the whole set of data.

AX Vel resulted to be the only DMC having the \fd 
amplitude larger than the \fu one. The rms residuals of the subsets are small
(0.007 and 0.010 mag for BA83 and PSB,  respectively) and  this allowed us
to detect the shallow  2\fu term. It is important to
note, once more, that also in the case of this small amplitude term the Fourier
decomposition supplies coherent and meaningful results: the BA83 dataset
yielded $\phi_{21}$=4.0$\pm$0.3 rad, while the PSB dataset yields 
$\phi_{21}$=4.4$\pm$0.2 rad.

 The comparison between the frequency values 
suggests to us a very stable behaviour of AX Vel. On the other hand, the 
amplitudes of the two frequencies are different in the two datasets. This
fact has two consequences: the rms residual of the whole set is slightly
higher than that of each dataset and some signal is left at the \fu and
\fd values in the final power spectrum. We cannot be sure that this
difference has a physical origin, since the measurements were collected in 
different photometric systems ($uvby, UBV, VBLUW$, ...) and instrumental
and/or transformation effects can originate the small (no more than 0.010 mag) 
discrepancies observed.

\subsection{AP Vel}
Similarly to AX Vel, 
we merged the three available subsets into two: the first was composed by
the LTPV data only, the second was formed by grouping the SB79 and P76
measurements. By comparing the results obtained with the two
different datasets, the very small differences found in the amplitude
values were considered as not significant.
The frequency analysis allowed us to detect the \fu, \fd,
\fm, \fp, 2\fu, 2\fd, 3\fu, 2\fu + \fd terms; the 2\fd+\fu and 3\fu+\fd terms
 were detected in one dataset and confirmed by the forced fit on the other.
On the other hand,
the small amplitude terms 4\fu,  4\fu + \fd were not
considered since the phase values were in disagreement. In the final
fit of the whole set of data (10 terms) we also added the BERD measurements.
\subsection{TU Cas}
The frequency analysis of the long term photometry of this DMC should give
an important answer about period variations. We separately analyzed three
large subsets (OO59, WE57, BERD),
spanning 50 years. The results do not support any trace of period variability:
we found, respectively,  0.46747, 0.46746, 0.46745 \cd for \fu (error bar:
$\pm$0.00001 \cd), 0.65859, 0.65860, 0.65864 \cd (error bar: $\pm$0.00002 \cd) 
for \fd. As regards the amplitudes, the values
determinated in the WE57 subset are appreciably larger than the
others (0.32 mag for \fu, 0.14 mag for \fd against 0.29 mag and 0.10,
respectively), but the physical meaning of this fact should be considered
with caution owing to instrumental differences. As a consequence, the
last power spectrum shows some residual signal around the \fu and \fd
values. The OO59 dataset was not
included in the global set owing to its large scatter (0.061 mag). Also the
Matthews et al. data (1992) could not be included, since its 
spectral window is very bad: this fact hampered a careful
frequency analysis and misleading results were obtained when forcing a fit.

The presence of a third period in the light curve of TU Cas is a controversial
point: false alarms are recurring in the DMC literature.
In agreement with the most recent results (Matthews et al. 1992),  we did not
find any trace of this third period.
\subsection{U TrA}
Six subsets were available (OO57, MI64, JA62, FS79, SB79, BERD) and they were
grouped
into three datasets. However, the dataset constituted by the MI64 and OO57
measurements showed a large scatter (more than 0.04 mag) and
it was used only to check the period values, which seem to be very stable;
only the \fu value obtained from the JA62 measurements (0.38947$\pm$0.00009
\cd) is marginally different from the other two (0.38934$\pm$0.00001 and 
0.38933$\pm$0.00002 \cd). As a consequence, the last power spectrum shows
some residual signal around the \fu value.
\subsection{EW Sct}
Figer et al. (1991; LTPV and Merate Observatory measurements) already reported
a Fourier  decomposition obtained by
using the procedure applied here to all the DMCs; in addition, we now
have at our disposal the 400 measurements collected by BERD.
The decomposition of the BERD data  allowed us to detect the
2\fd, 2\fu + \fd and 3\fu terms, whose phase values were confirmed by the
forced fit on Figer et al.'s data. However, a small difference in the
amplitude of the \fu and \fd terms is observed when comparing the FP91
and BERD subsets; the difference is particularly significant for the \fd term
(0.114 mag in the FP91 data, 0.127 in the BERD one) and may suggest a 
physical variation. As a consequence,
the last power spectrum  shows some residual signal at values close to the
\fd value.
\subsection{VX Pup}
The frequency analysis of  the datasets obtained by merging the SB79
and MB84 measurements and considering the LTPV data only are
very similar; amplitude and frequency values are within the error bars.
The \fu, \fd, \fm, \fp, 2\fu, 2\fd, 3\fu, 2\fu + \fd, 2\fd + \fu are detected
in both datasets; the 2\fd--\fu  term only in the LTPV dataset, but its
reality was confirmed. The phases of the 3\fu+\fd term are only marginally
coincident, but the power spectrum obtained by introducing all the above terms
clearly showed it and therefore it was considered in the global fit, 
which also considered the BERD measurements. It should be noted that
the two close terms \fd--\fu=0.1353 \cd and 2\fu--\fd=0.1314 \cd are both
observed in the power spectra.
\subsection{Y Car}
The available data were firstly  grouped into two datasets (SB79+P76, BERD).
The frequency analysis allows us to detect the \fu, \fd, \fm, \fp,
2\fu, 2\fd, 3\fu, 2\fu--\fd, 2\fd+\fu, 3\fu+\fd. SB79 only considered the first
six terms: the identification of the other terms allowed us to reduce the rms
residual from 0.024 mag to 0.017 mag. When forming the whole set of data, the
rms again increased to 0.021 mag, owing to the lesser accuracy of the BERD
data. The last power spectrum is very noisy and a peak at 0.94 \cd (or 1.06
\cd) is visible; since the two coupling terms 0.940182 \cd (2\fu+\fd) and
1.056138\cd (\fu+2\fd) were already considered, its nature is not obvious.
However, the number of measurements is quite small and these
results can be an artifact due to poor sampling.
\subsection{AS Cas}
The photoelectric measurements carried out by BERD do not have the same
mean magnitude from one year to the next; first, they were aligned to the
same value (maximum correction: 0.086 mag) before performing the frequency
analysis described in Sect. 2. In spite of this, a spurious peak was detected
at 1.002 \cd. The rms residual (0.028 mag)
is high (in particular the measurements reported by Berdnikov et al. 1995
display a large scatter), but it should be noticed that AS Cas has a mean
magnitude  $V$=12.26
and that it was observed with a 60--cm reflector.
\subsection{BK Cen}
The precision of the measurements in the available subsets (LJ65, SB79,
LTPV and BERD) is different. We analyzed them separately and
we found very similar amplitude and phase values, confirming
the internal stability of the proposed solution.
In particular, the frequency analysis of the
LJ65 and LTPV subsets yielded  \fu and \fd values coincident within error
bars, suggesting no period variation over more than 30 years.
The rms residual (0.0231 mag) and the residual
noise amplitude are rather high (0.004 mag). This can be due to a high
number of measurements with a residual between 3$\sigma$ and 4$\sigma$ which
we preferred not to delete. 
\subsection{V367 Sct}
To perform photometric measurements of this faint DMC belonging to the open 
cluster NGC 6649 is not an easy task, but the last 
power spectrum did not show any residual high level peak. Even if large error
bars prevent a detailed analysis, frequency and amplitude values seems to be
stable in the three available subsets (MS78, MV75, BERD): the \fu, \fd, \fm,
\fp, 2\fu, 2\fd terms are detected in all the subsets.
The small amplitude 3\fu term was firstly detected in the BERD subset and then
confirmed by the fit on the other subsets.
\subsection{GZ Car}
The frequency analysis was carried out by combining the SB79 and P76 subsets;
despite the small number of measurements (91), the harmonics 2$f$ of both
frequencies, the coupling terms \fm and \fp and the 2\fu + \fd were found.
The detection of the small amplitude 2\fd term is notable. To have a
solution based on a higher number of points, the BERD subset was added.
\subsection{UZ Cen}
We performed the frequency analysis by combining the SB79 and P76 subsets.
As in the case of AP Vel,
the BERD subset was added to perform the global fit. The \fu light curve
is very asymmetrical (3 harmonics are required); on the other hand, the \fd
amplitude is the smallest observed in the whole sample and the shallow 
2\fd term was detected only in the last steps of the analysis.  

\subsection{CO Aur}
The three subsets (AM84+AM86, BB87, BERD; in the latter the annual 
misalignment was corrected) 
yield very similar results for amplitude and frequency values. It is
important to note that the 2\fd term cannot be detected in a reliable way
and hence the light curve on \fd must be considered to be sine--shaped.
In addition to \fu, \fd, 2\fu, the terms \fp, \fm, 3\fu can be detected.
\begin{table*}
\begin{flushleft}
\caption{Coefficients of the least--squares fits of the whole sets of 
data: TU Cas, U TrA,
UZ Cen, AS Cas, VX Pup, BK Cen, AP Vel, Y Car, EW Sct }
\begin{tabular}{l | rrr | rrr | rrr}
\hline
\multicolumn{1}{c}{Term}  &  \multicolumn{1}{|c}{Frequency} & \multicolumn{1}{c}{Ampl.} & \multicolumn{1}{c|}{Phase}  &
                             \multicolumn{1}{|c}{Frequency} & \multicolumn{1}{c}{Ampl.} & \multicolumn{1}{c|}{Phase}  &
                             \multicolumn{1}{c}{Frequency} & \multicolumn{1}{c}{Ampl.} & \multicolumn{1}{c}{Phase} \\
\multicolumn{1}{c}{} &  \multicolumn{1}{|c}{[cd$^{-1}$]} & \multicolumn{1}{c}{[mmag]} & \multicolumn{1}{c|}{[rad~10$^{-2}$]} & 
                         \multicolumn{1}{c}{[cd$^{-1}$]} & \multicolumn{1}{c}{[mmag]} & \multicolumn{1}{c|}{[rad~10$^{-2}$]} & 
                         \multicolumn{1}{c}{[cd$^{-1}$]} & \multicolumn{1}{c}{[mmag]} & \multicolumn{1}{c}{[rad~10$^{-2}$]} \\
\noalign{\smallskip}
\hline
\hline
\noalign{\smallskip}
     & \multicolumn{3}{c}{TU Cas} & \multicolumn{3}{|c}{U TrA} & \multicolumn{3}{|c}{UZ Cen}\\
\noalign{\smallskip}
\hline
\noalign{\smallskip}
 \fu       & 0.467442 & 292\t1 & 431\t1&  0.389344 & 263\t1 & 520\t1 &  0.299910 & 291\t3 & 16\t1 \\  
 \fd       & 0.658635 & 114\t1 & 219\t2 &  0.547983 & 101\t1 &625\t1 &  0.424589 & 82\t3  &623\t4 \\  
2\fu       &          & 101\t1 & 21\t1 &           & 85\t1 &  199\t1 &           & 95\t3 &454\t4 \\
\fu+\fd    &          & 78\t1 &413\t1 &           & 65\t1 &  293\t1&           & 54\t3 &428\t6 \\
\fd-\fu    &          & 37\t1 & 192\t3 &           & 26\t1 & 479\t4 &           & 21\t3 &405\t15 \\
2\fd       &          & 14\t1 & 243\t8 &           &10\t2  &433\t8  &           & 10\t3 &439\t31 \\
3\fu       &          & 37\t1 & 244\t3 &           & 28\t1 &512\t3 &           & 41\t3 & 267\t7 \\         
2\fu+\fd   &          & 49\t1 & 14\t2 &           & 46\t2 & 601\t2 &           &32\t3 & 255\t9 \\
\fu+2\fd   &          & 19\t1 & 419\t6 &           & 15\t1 &  67\t6 &           &  & \\
2\fd-\fu   &          & 7\t1 & 172\t16 &           &  &   &           &  &  \\
4\fu       &          & 12\t1 & 444\t9 &           & 9\t1 &  156\t9 &           & 14\t3 & 82\t22 \\
3\fu+\fd   &          & 23\t1 & 236\t4 &           & 19\t2 &  296\t4 &           & 22\t3 & 52\t14 \\
2\fu+2\fd   &          & 15\t1 & 2\t7 &           &  &   &           &  &  \\
3\fu-\fd   &          &       &       &                &  &       &           &7\t3  &105\t45  \\
3\fu+2\fd  &          & 7\t1 & 211\t14 &           &  &   &           &  &  \\
4\fu+\fd   &          & 11\t1 & 445\t9 &           & 9\t1 & 600\t10 &           &  &  \\
\noalign{\smallskip}
$V_0$     & \multicolumn{3}{|c|}{7.7687$\pm$0.0007} & \multicolumn{3}{|c|}{7.9695$\pm$0.0006} & \multicolumn{3}{|c}{8.8000$\pm$0.0018}\\
rms       & \multicolumn{3}{|c|}{0.0161 mag} & \multicolumn{3}{|c|}{0.0167 mag} & \multicolumn{3}{|c}{0.0190 mag}\\
N         & \multicolumn{3}{|c|}{618} & \multicolumn{3}{|c|}{1060} & \multicolumn{3}{|c}{131}\\
$T_0$     & \multicolumn{3}{|c|}{HJD 2448752.3129} & \multicolumn{3}{|c|}{HJD 2436764.8808} & \multicolumn{3}{|c}{HJD 2442125.1938}\\
\noalign{\smallskip}
\hline
\hline
\noalign{\smallskip}
     & \multicolumn{3}{c}{AS Cas} & \multicolumn{3}{|c}{VX Pup} & \multicolumn{3}{|c}{BK Cen}\\
\noalign{\smallskip}
\hline
\noalign{\smallskip}
 \fu       & 0.330628  &   203\t2  &   90\t1 &  0.332030 & 174\t1 & 186\t1 &  0.315072 & 266\t2 &  39\t1 \\  
 \fd       & 0.463936  &   137\t2  &  242\t1 &  0.467384 & 144\t1 & 408\t1 &  0.449860 & 108\t2 & 116\t2 \\
 2\fu      &           &    62\t2  &  595\t3 &           & 32\t1 & 158\t6 &           &  70\t2 & 501\t3 \\ 
 \fp       &           &    67\t2  &  116\t2 &           &  51\t1 &380\t3 &           &  54\t2 & 582\t4 \\
 \fm       &           &    35\t2  &  561\t5 &           &  25\t1 &  44\t7 &           &  26\t2 &521\t8 \\
 2\fd      &           &    19\t2  &  290\t9 &           &  17\t1 &  7\t11 &           &  11\t2 &  92\t19\\
 3\fu      &           &    16\t2  &  489\t12&           &   7\t1 & 161\t30&           &  26\t2 &325\t8 \\
2\fu+\fd   &           &    36\t2  &    4\t5 &           &  19\t1 & 358\t10 &           &  32\t2 &408\t6 \\
\fu+2\fd   &           &    19\t2  &  189\t9 &           &  12\t1 & 602\t14&           &  11\t2 &500\t18\\
2\fd-\fu   &           &    11\t2  & 613\t15 &           &   7\t1 & 240\t22&                             \\
3\fu+\fd   &           &    16\t2  & 513\t11&           &   5\t1 & 378\t30&           &  11\t2 & 226\t19\\
\noalign{\smallskip}
$V_0$     & \multicolumn{3}{|c|}{12.2689$\pm$0.0015} & \multicolumn{3}{|c|}{8.3201$\pm$0.0008} & \multicolumn{3}{|c}{10.1333$\pm$0.0015}\\
rms       & \multicolumn{3}{|c|}{0.0265 mag} & \multicolumn{3}{|c|}{0.0121 mag} & \multicolumn{3}{|c}{0.0231 mag}\\
N         & \multicolumn{3}{|c|}{575} & \multicolumn{3}{|c|}{234} & \multicolumn{3}{|c}{251}\\
$T_0$     & \multicolumn{3}{|c|}{HJD 2448648.4725} & \multicolumn{3}{|c|}{HJD 2443803.1587} & \multicolumn{3}{|c}{HJD 2446319.7188}\\
\noalign{\smallskip}
\hline
\hline
\noalign{\smallskip}
 & \multicolumn{3}{c}{AP Vel} & \multicolumn{3}{|c}{Y Car} & \multicolumn{3}{|c}{EW Sct}\\
\noalign{\smallskip}
\hline
\noalign{\smallskip}
 \fu       & 0.319717  &   279\t1  & 334\t1 &  0.274742 & 265\t3 & 311\t1 &  0.171719 & 171\t1 & 432\t1 \\  
 \fd       & 0.454587  &   137\t1  & 485\t1 &  0.390698 & 117\t3 & 518\t2 &  0.245820 & 124\t2 & 212\t1 \\
 2\fu      &           &    79\t1  & 457\t2 &           &  79\t3 & 412\t4 &           &  28\t2 &  48\t6 \\ 
 \fp       &           &    52\t2  & 587\t3 &           &  68\t3 & 628\t4 &           &  31\t1 & 485\t6 \\
 \fm       &           &    36\t1  & 592\t4 &           &  32\t3 &  37\t10 &           & 20\t2 & 249\t8 \\
 2\fd      &           &    16\t2  &  159\t11&           &  12\t3 & 213\t22 &           &  5\t2 &  96\t32\\
 3\fu      &           &    26\t2  & 604\t6  &           &  28\t3 & 531\t10&           &  40\t1 & 295\t50 \\
2\fu+\fd   &           &    28\t2  &  509\t6 &           &  21\t3 & 108\t13 &           &  57\t2  &54\t30 \\
\fu+2\fd   &           &    12\t2  &  250\t14&           &   9\t3 & 318\t31&           &   & \\
3\fu+\fd   &           &     6\t1  &  191\t23&           &  14\t3 & 205\t21&           &   & \\
\noalign{\smallskip}
$V_0$     & \multicolumn{3}{|c|}{10.0610$\pm$0.0010} & \multicolumn{3}{|c|}{8.1080$\pm$0.0020} & \multicolumn{3}{|c}{7.9888$\pm$0.0007}\\
rms       & \multicolumn{3}{|c|}{0.0158 mag} & \multicolumn{3}{|c|}{0.0213 mag} & \multicolumn{3}{|c}{0.0165 mag}\\
N         & \multicolumn{3}{|c|}{255} & \multicolumn{3}{|c|}{137} & \multicolumn{3}{|c}{515}\\
$T_0$     & \multicolumn{3}{|c|}{HJD 2446298.5843} & \multicolumn{3}{|c|}{HJD 2442250.5989} & \multicolumn{3}{|c}{HJD 2446300.2641}\\
\noalign{\smallskip}
\hline
\hline
\noalign{\smallskip}
\end{tabular}
\end{flushleft}
\end{table*} 
\begin{table*}
\begin{flushleft}
\caption{Coefficients of the least--squares fits of the whole sets of 
data: AX Vel, GZ Car, V367 Sct, CO Aur. For BQ Ser see the last panel of
Tab. 3}
\begin{tabular}{l | rrr | rrr}
\hline
\multicolumn{1}{c}{Term}  &  \multicolumn{1}{|c}{Frequency} & \multicolumn{1}{c}{Ampl.} & \multicolumn{1}{c|}{Phase}  &
                             \multicolumn{1}{|c}{Frequency} & \multicolumn{1}{c}{Ampl.} & \multicolumn{1}{c}{Phase}  \\
\multicolumn{1}{c}{} &  \multicolumn{1}{|c}{[cd$^{-1}$]} & \multicolumn{1}{c}{[mmag]} & \multicolumn{1}{c|}{[rad~10$^{-2}$]} & 
                         \multicolumn{1}{c}{[cd$^{-1}$]} & \multicolumn{1}{c}{[mmag]} & \multicolumn{1}{c}{[rad~10$^{-2}$]}\\
\noalign{\smallskip}
\hline
\hline
\noalign{\smallskip}
     & \multicolumn{3}{c}{AX Vel} & \multicolumn{3}{|c}{GZ Car} \\
     \noalign{\smallskip}
\hline
\noalign{\smallskip}
 \fu       & 0.272241  &   106\t1  &   23\t1 &  0.240448 & 147\t2 & 265\t1  \\  
 \fd       & 0.385657  &   143\t1  &  469\t1 &  0.390698 &  87\t2 &  62\t2  \\
 2\fu      &           &    11\t2  &  463\t7 &           &  23\t2 & 335\t8  \\ 
 \fp       &           &    28\t1  &  300\t3 &           &  25\t2 & 152\t7  \\
 \fm       &           &    12\t1  &  284\t7 &           &  16\t2 & 225\t11 \\
 2\fd      &           &    11\t1  &  138\t8 &           &   4\t2 & 594\t45\\
2\fu+\fd   &           &     6\t1  &  170\t15 &           &    &           \\
2\fd-\fu   &           &           &          &          &   8\t2 & 24\t22 \\
\noalign{\smallskip}
$V_0$     & \multicolumn{3}{|c|}{8.2148$\pm$0.0007} & \multicolumn{3}{|c}{10.2387$\pm$0.0013}\\
rms       & \multicolumn{3}{|c|}{0.0109 mag} & \multicolumn{3}{|c}{0.0130 mag} \\
N         & \multicolumn{3}{|c|}{520} & \multicolumn{3}{|c}{118} \\
$T_0$     & \multicolumn{3}{|c|}{HJD 2443892.0164} & \multicolumn{3}{|c}{HJD 2442073.2970}\\
\noalign{\smallskip}
\hline
\hline
\noalign{\smallskip}
     & \multicolumn{3}{c}{V367 Sct} & \multicolumn{3}{|c}{CO Aur} \\
     \noalign{\smallskip}
\hline
\noalign{\smallskip}
 \fu       & 0.158902  &   176\t1  & 542\t1 &  0.560844 & 173\t1 & 288\t1  \\  
 \fd       & 0.228061  &   117\t1  & 203\t1 &  0.700390 &  43\t2 & 233\t4  \\
 2\fu      &           &    27\t1  & 278\t5 &           &  31\t1 & 357\t6  \\ 
 \fp       &           &    18\t1  & 519\t7 &           &  9\t1 &  345\t19  \\
 \fm       &           &    15\t1  &  160\t9 &           &  7\t2 & 418\t23 \\
 2\fd      &           &    14\t1  &  140\t10 &           &      &          \\
3\fu       &           &     4\t1  &  588\t33 &           & 7\t2 & 419\t26  \\
\noalign{\smallskip}
$V_0$     & \multicolumn{3}{|c|}{11.6073$\pm$0.0009} & \multicolumn{3}{|c}{7.7142$\pm$0.0008}\\
rms       & \multicolumn{3}{|c|}{0.0207 mag} & \multicolumn{3}{|c}{0.0146 mag} \\
N         & \multicolumn{3}{|c|}{514} & \multicolumn{3}{|c}{370} \\
$T_0$     & \multicolumn{3}{|c|}{HJD 2448093.7156} & \multicolumn{3}{|c}{HJD 2445758.3582}\\
\noalign{\smallskip}
\hline
\hline
\noalign{\smallskip}
\end{tabular}
\end{flushleft}
\end{table*} 
\begin{figure*}[t]
\epsfxsize=18truecm
\centerline{\epsffile[25 430 550 710]{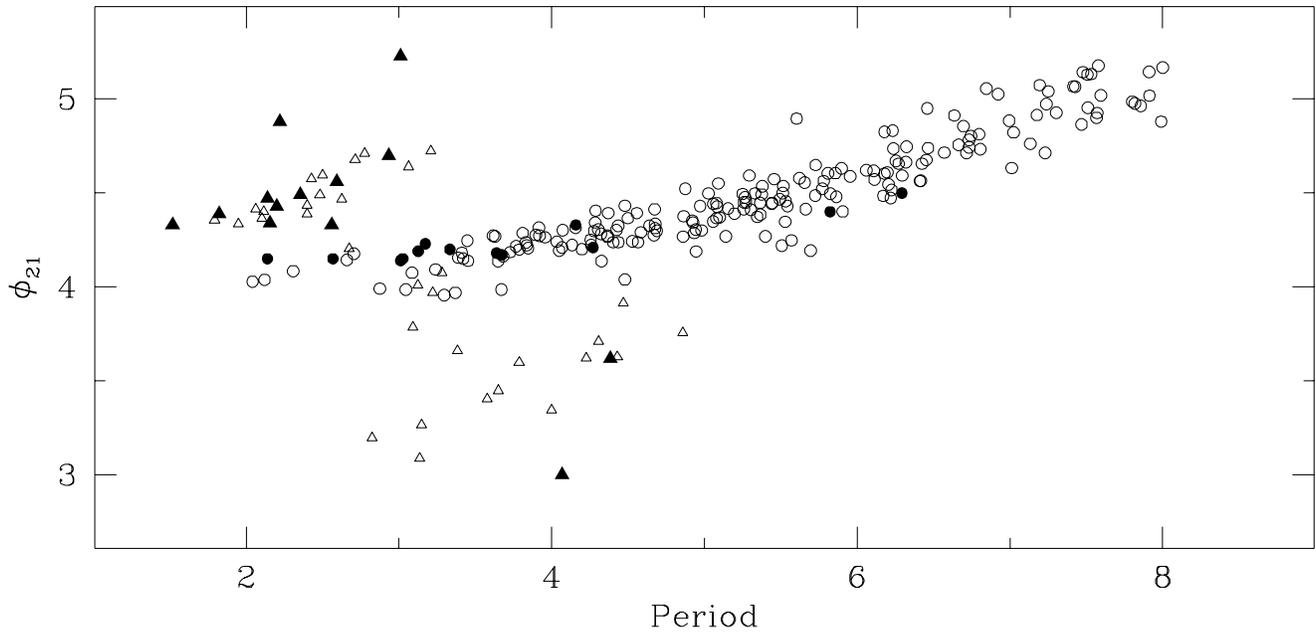}}
\caption[]{The $\phi_{21}$--$P$ plane. Dots: single--mode Classical
Cepheids. Triangles: $s$--Cepheids. Filled dots: Fundamental radial mode
of DMCs. Filled triangles: 1$O$ radial mode of DMCs.}
 \end{figure*}
\section{Discussion and Conclusions}
The frequency analysis performed by the least--squares method
allowed us to obtain a very detailed description of the light curves of the
galactic DMCs. With respect to the goals of this first investigation, 
some conclusions can be extracted directly from the analysis 
reported in the previous section:
\begin{enumerate}
\item The 2$^{\rm nd}$ order terms are present in the light curves of all 
the stars, but in every case  a fit limited to the 2$^{\rm nd}$ order is
not satisfactory.
As regards the \fu component, only AX Vel and GZ Car do not show the
3\fu harmonic, while TU Cas, U TrA and UZ Cen show also the 4$^{\rm th}$
harmonic. As regards the \fd component, only the first harmonic is observed 
in its light curve. The coupling terms are observed in a large variety of
combinations. The \fp and \fm terms are observed in all the stars and also the
2\fu+\fd term is rather common. Curiously, Alcock et al. (1995) and Welch et al.
(1996) presented
only the 2$^{\rm nd}$ order components in their discussion of DMC light
curves in the LMC; probably a deeper analysis can yield some other interesting
results.
\item The two independent frequencies \fu and \fd seem to be very stable, in
the sense that a reasonable fit can be obtained without admitting their
variation. U TrA is the most promising candidate to show such a variation,
since a slightly different \fu value was obtained for the oldest subset;
\item In none of the stars a convincing third independent periodicity is detected,
even in the cases of TU Cas and BQ Ser, the two claimed candidates;
\item The amplitudes of the modes do not show variations 
exceeding the error bars, with the exception of the \fd term in the EW
Sct light curve;  this star it is  the most suitable target for an extensive
long-term photometry project carried out by using a very stable 
instrumentation.
 Berdnikov (1992) showed how the light curve changes in amplitude over
a period when considering different phases of the other
period. However, this effect is not real, since it is due to the presence of
the cross coupling terms, which Berdnikov did not subtract from the
original measurements; when considering
these terms,  light curves with constant amplitude over each period can
be easily constructed, as Fig. 3 shows for BQ Ser. A full set of light curves
over the two periods for each DMC can be found in Pardo (1995).
\end{enumerate}
In Introduction we mentioned the separation between Classical and $s$--Cepheids
in the space of Fourier parameters; Antonello et al. (1990) ascribed this 
separation to the different pulsation mode and also invoked the action of
a resonance at or near 3.0 d to explain the ``$Z$" shape of the $s$--Cepheid
progression.  The very reliable Fourier parameters now at our disposal
for the galactic DMCs allow us to give an independent confirmation of these
intrepretations. Figure 4 shows the distribution of the $\phi_{21}$ values of
the galactic DMCs superimposed to the Classical and {\it s}--Cepheids
ones. The $\phi_{21}$ values corresponding to the $F$ radial mode occupy the
same region as the Classical Cepheids. In like manner, the $\phi_{21}$ values
of the the 1$O$ radial mode mimics the ``$Z$" shape: note the overlap between
DMCs and {\it s}--Cepheids in the upper part, the high value at
3.0 d (BQ Ser) and the positioning of the two $\phi_{21}$ values belonging to
the longest period DMCs (EW and V367 Sct) just on the lower part.
It appears  quite evident that in the DMCs the light curves of the $F$--radial
mode and the 1$O$--mode are very similar to the curves of the Classical and
$s$--Cepheids, respectively. In turn, this fact proves without any doubt
that $s$--Cepheids are pulsating in the 1$O$ mode and that the $\phi_{21}$
value can be considered a powerful discriminant between these modes. 
It should be also noted that the $F$--mode light curve follows the
Hertzsprung progression. A discontinuity is present near 3.0 d in the
light curves of 1$O$ pulsators and a resonance effect is the more likely
cause. 

The case of CO Aur deserves a particular attention. The ratio between the
observed frequencies is 0.800 and this value is explained by the excitation
of the 1$O$ and 2$O$ modes. In the $\phi_{21}$--$P$ plane the $\phi_{21}$
value for the \fu term falls in the short period region, where the 1$O$
and $F$ sequencies are merging; we can only conclude that the $\phi_{21}$
value for this unique (in the Galaxy) pulsator is quite similar to the
others. It has not been possible to detect the 2\fd term, i.e. the \fd
light curve is perfectly sine--shaped. Stellingwerf et al. (1987) predicted
an asymmetrical light curve for a 2$O$ pulsator, but this does not seem to be
verified in the CO Aur case. It should also be noted that between the single--mode
Cepheids there are two stars (V1334 Cyg and DT Cyg) showing a perfectly 
sine--shaped
light curve (Poretti 1994). In view of the close similarity evidenced above
between single and double--mode pulsators, further investigation of the
pulsating mode of V1334 Cyg and DT Cyg is recommended.

\begin{acknowledgements} The authors wish to thank E.~Antonello,  
L.~Mantegazza, L.~Pasinetti for useful discussions and
J. Vialle for the improvement of the English form of the manuscript.
\end{acknowledgements}

\end{document}